\documentclass[a4paper]{jpconf}
\usepackage{graphicx}

\usepackage{subfigure}

\usepackage{amsmath}

\usepackage{cite}

\usepackage{placeins}

\usepackage[hyperindex,breaklinks]{hyperref}

\def\ttbar {\mbox {$t\bar{t}$}}

\def\pT{\mbox{$p_{\text{T}}$}}

\def\MET{\ensuremath{E_{\mathrm{T}}^{\mathrm{miss}}}}

\def\MTW{\mbox{$m_\mathrm{T}(\ell , \MET)$}}
\def\mfmet{\ensuremath{m(f_{\mathrm{met}})}}
\def\mvmet{\ensuremath{m(v_{\mathrm{met}})}}
\def\vmet{\ensuremath{v_{\mathrm{met}}}}
\def\fmet{\ensuremath{f_{\mathrm{met}}}}
\def\ares{\ensuremath{a_{\mathrm{res}}}}
\def\anonres{\ensuremath{a_{\mathrm{non\mbox{-}res}}}}
\def\deltaphilb{\ensuremath{\Delta\phi(\ell,b)}}
\def\adeltaphilb{\ensuremath{|\Delta\phi(\ell,b)|}}

\DeclareGraphicsExtensions{.pdf,.png}

\begin{document}
\title{Search for monotop events using the ATLAS detector at the LHC}

\author{Timoth\'ee Theveneaux-Pelzer\\\textit{on behalf of the} ATLAS \textit{collaboration}}

\ead{tpelzer@cern.ch}

\begin{abstract}
A search for the production of single-top-quarks in association with missing energy performed in
proton--proton collisions at a centre-of-mass energy of $\sqrt{s}=\mathrm{8~TeV}$ with the ATLAS experiment is presented.
No deviation from the Standard Model prediction is observed, and upper limits are set on the production cross-section
for resonant and non-resonant production of an invisible exotic state in association with a right-handed top quark.
\end{abstract}

\section{Introduction}

Several theories beyond the Standard Model (BSM) predict events -- dubbed ``monotop'' -- where a single-top-quark is produced in association with
large missing energy.
A search for such events in the proton--proton collisions produced in 2012 by the Large Hadron Collider
at a centre-of-mass energy of $\sqrt{s}=\mathrm{8~TeV}$ has been performed by the ATLAS experiment,
using data corresponding to an integrated luminosity of $20.3$~fb$^{-1}$~\cite{ATLASMonotop}.
In this search, the $W$ boson from the top quark is required to decay into an electron or a muon and a neutrino.

\section{Signal models}

Monotop events appear in a wide range of BSM theories, therefore a theory--independent approach is followed.
Following Refs.~\cite{AndreaFuksMaltoni,Wangetal,Cacciapagliaetal}, two effective models are used:
\begin{itemize}
 \item Resonant production of a +2/3 charged, spin-0 boson $S$ with a mass of 500~GeV, decaying into a right-handed top quark
 and a neutral, colour singlet, spin-1/2 fermion, $\fmet$, with a mass between 0 and 100~GeV;
 \item Non-resonant production of a neutral, colour singlet, spin-1 boson, $\vmet$, with a mass between 0 and 1000~GeV, in association with a right-handed top quark.
\end{itemize}

\begin{figure}[!h!tb]
\centering
\subfigure[\label{subfig:feyn_Res}]{\includegraphics[width=0.3\textwidth]{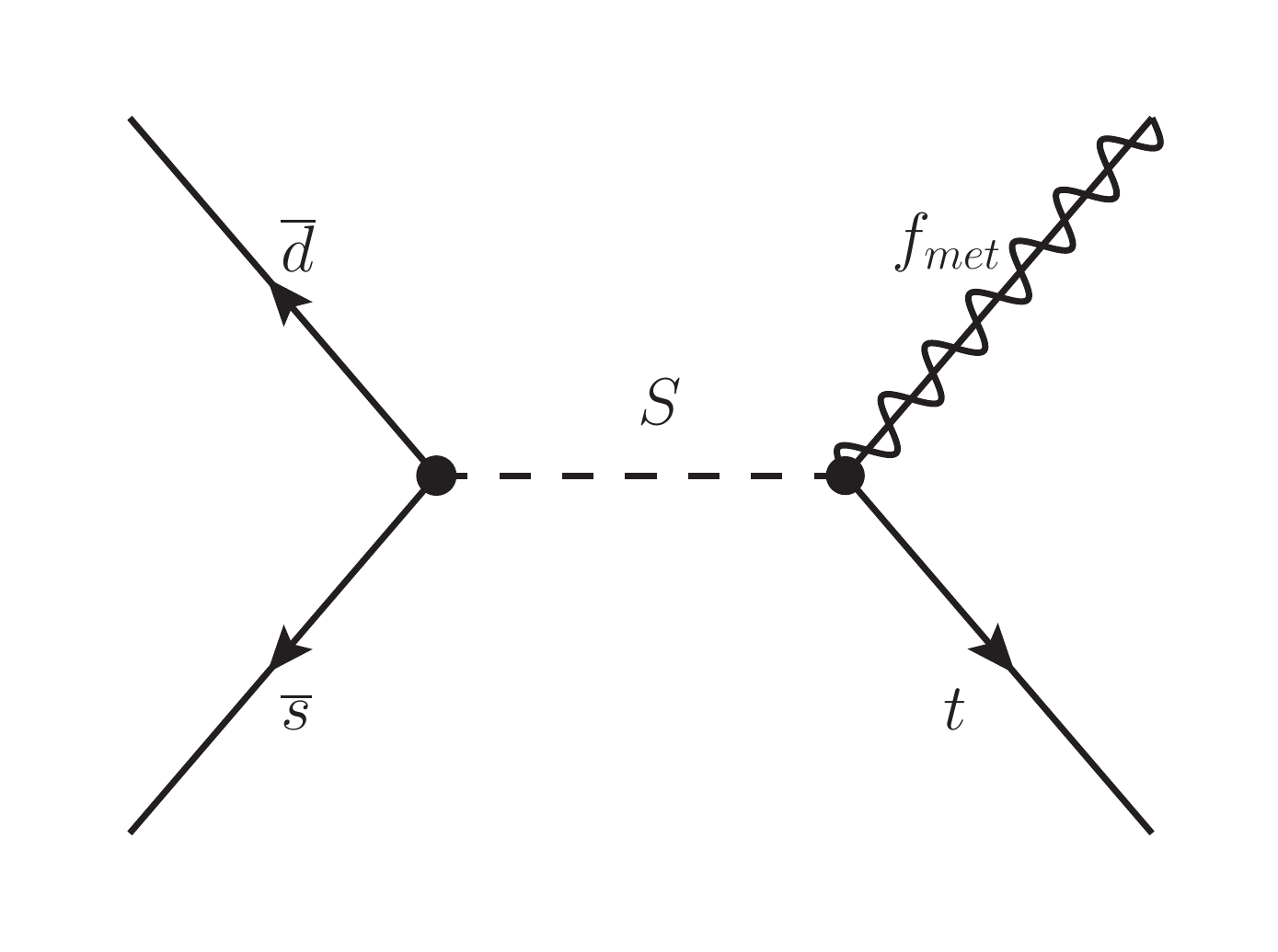}}
\subfigure[\label{subfig:feyn_NonRes_schan}]{\includegraphics[width=0.3\textwidth]{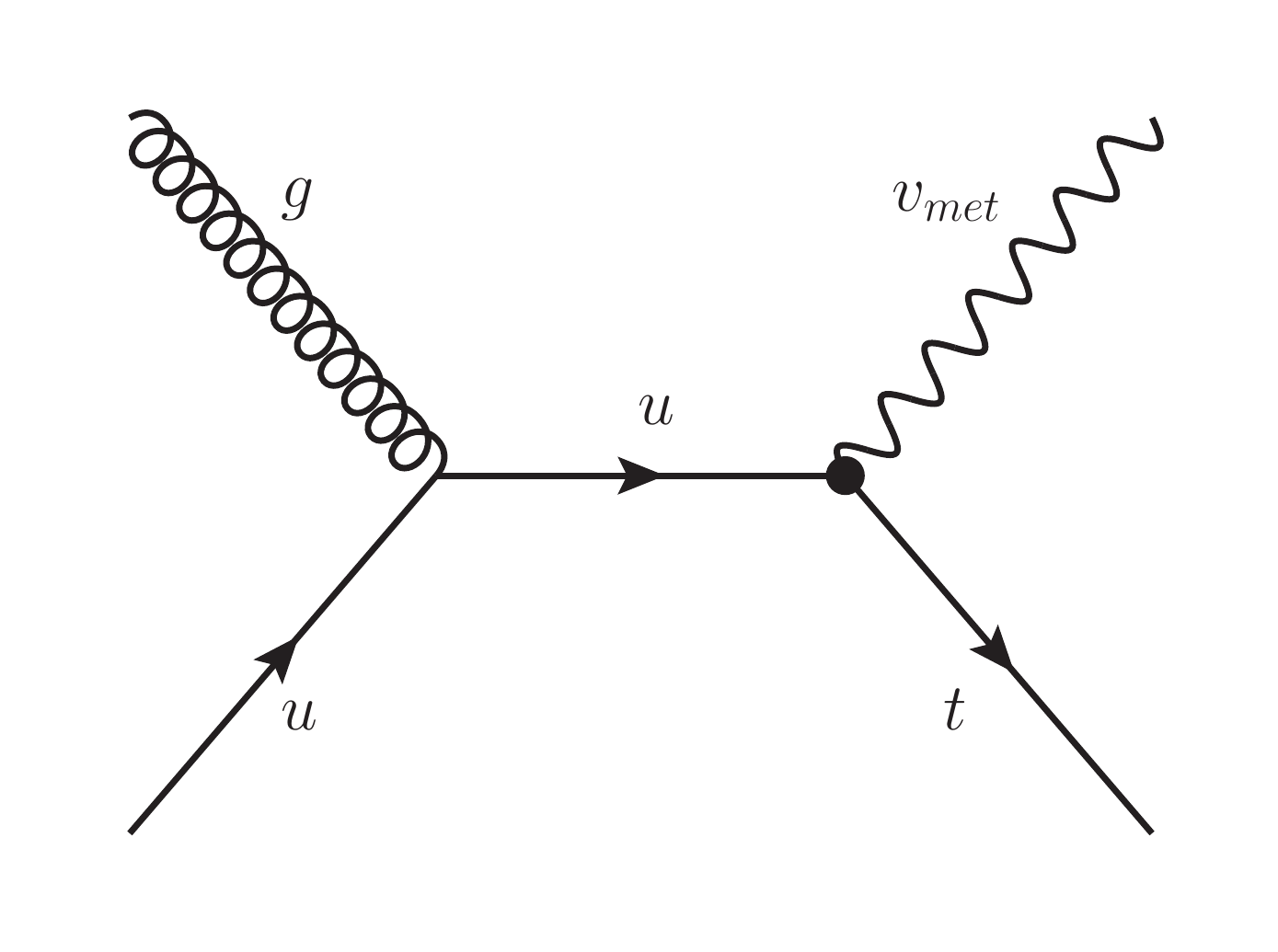}}
\subfigure[\label{subfig:feyn_NonRes_tchan}]{\includegraphics[width=0.3\textwidth]{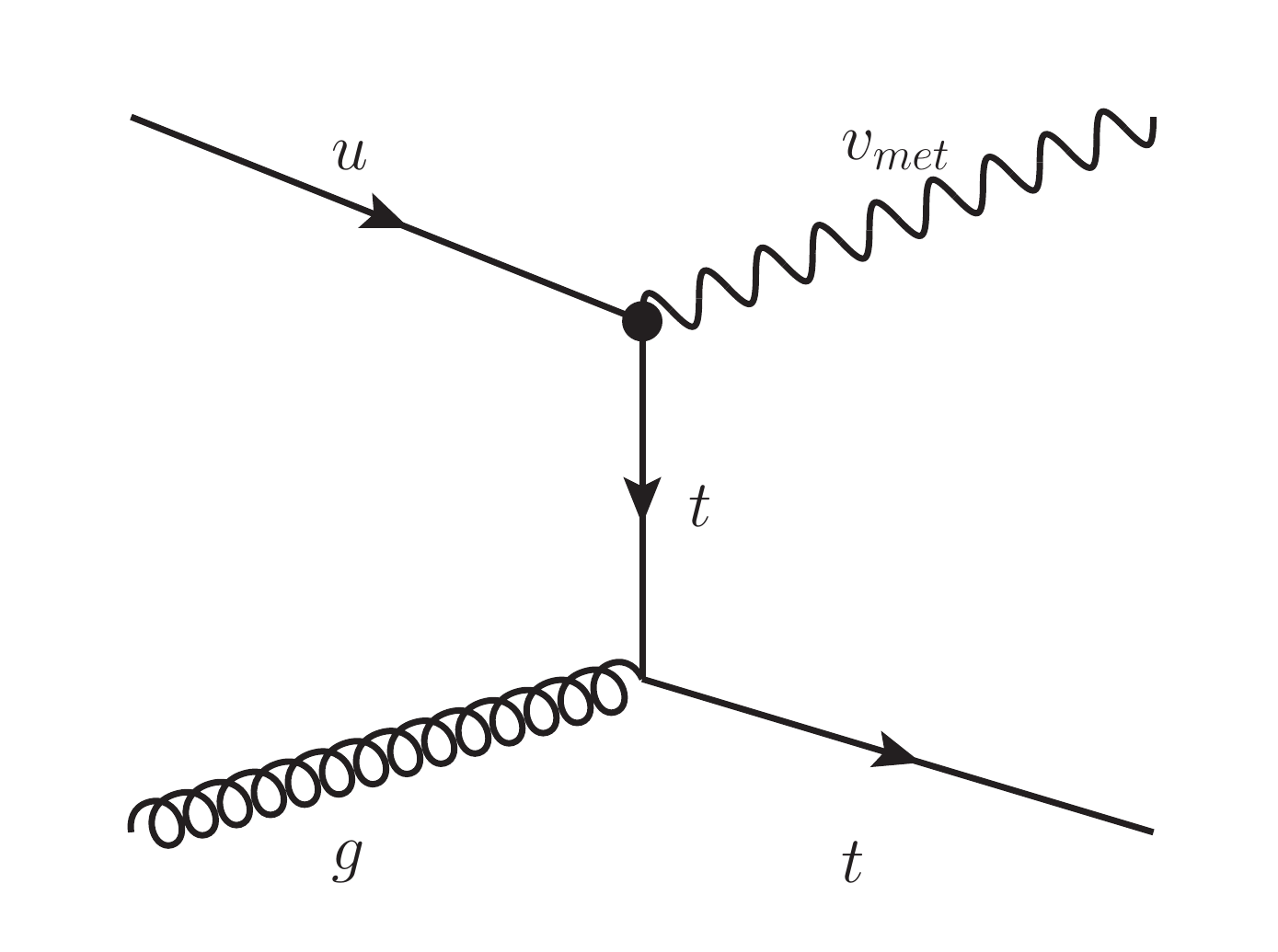}}
\caption
{
Example of Feynman diagrams of leading-order processes leading to monotop events: \subref{subfig:feyn_Res}~production of
a coloured scalar resonance $S$ decaying into a top quark and a spin-$1/2$ fermion $\fmet$
in the resonant model, and \subref{subfig:feyn_NonRes_schan}~$s$-
and \subref{subfig:feyn_NonRes_tchan}~$t$-channel non-resonant production of a top quark in association with
a spin-1 boson $\vmet$ in the non-resonant model.
}
\label{fig:feyn}
\end{figure}

Figure~\ref{fig:feyn} shows Feynman diagrams corresponding to these two effective models.
The interaction of the BSM particles with the SM fermions are characterised by coupling strengths,
denoted \ares\ and \anonres\ for the resonant and non-resonant models, respectively,
and are independent of the masses of the new particles.
The models have been implemented in {\sc FeynRules}, and simulated signal events
have been generated using the generator {\sc MadGraph5} interfaced with {\sc Pythia8}.

\section{Analysis strategy}

The experimental signature consists of one prompt electron or muon, one $b$-tagged jet and large missing transverse momentum.
The main background processes are \ttbar\ production -- mostly due to dileptonic decays for which one prompt lepton and one $b$-jet
fail the selection criteria -- and $W$+jets events.
Sub-dominant sources of background events arise from single-top,  $Z$+jets, diboson and multijet production.
The expected contributions of background events are estimated with simulated event samples,
except the multijet production which is estimated with a data-driven matrix-method.

A cut--and--count approach is followed.
Events with one prompt electron or muon and one jet are selected.
The jet is required to be identified as originating from a $b$-quark ($b$-tagged).
The multijet background is rejected by imposing the magnitude of the missing transverse momentum \MET\ to be larger that 35~GeV,
and the transverse mass\footnote{The transverse mass is defined as\\$\MTW=\sqrt{2 \pT\left(\ell\right) \MET \left(1-\cos \Delta \phi \left(\pT\left(\ell\right), \MET\right) \right)}$,
where $\pT\left(\ell\right)$ denotes the modulus of the lepton transverse momentum, and $\Delta \phi \left(\pT\left(\ell\right), \MET\right)$
the azimuthal difference between the missing transverse momentum and the lepton directions.}
of the lepton-\MET\ system \MTW\ is required to be larger that 60~GeV.

\begin{figure}[!htb]
\centering
\subfigure[\label{subfig:BkgSig_MtW}]{\includegraphics[width=0.42\textwidth]{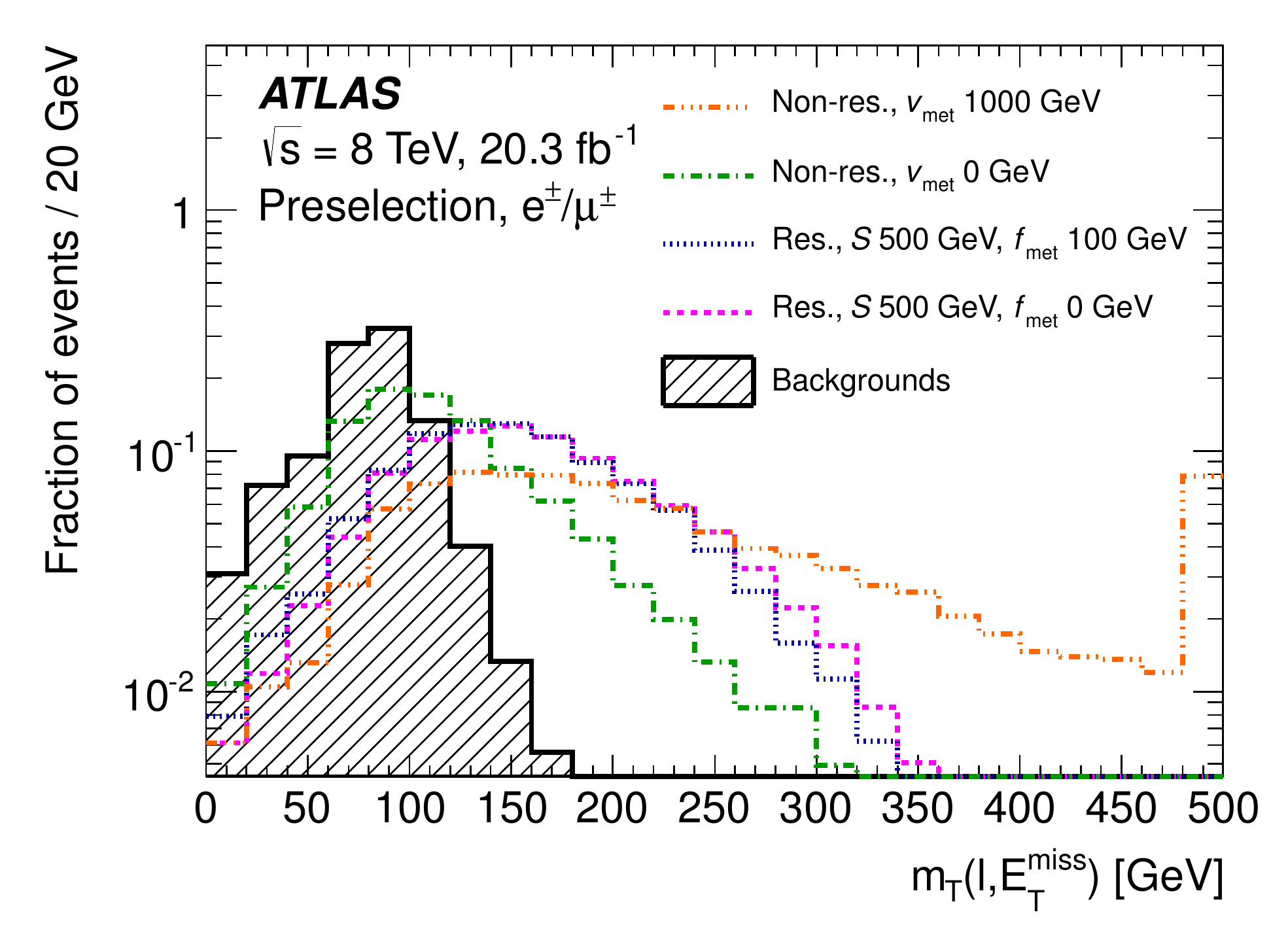}}
\subfigure[\label{subfig:BkgSig_DPhi}]{\includegraphics[width=0.42\textwidth]{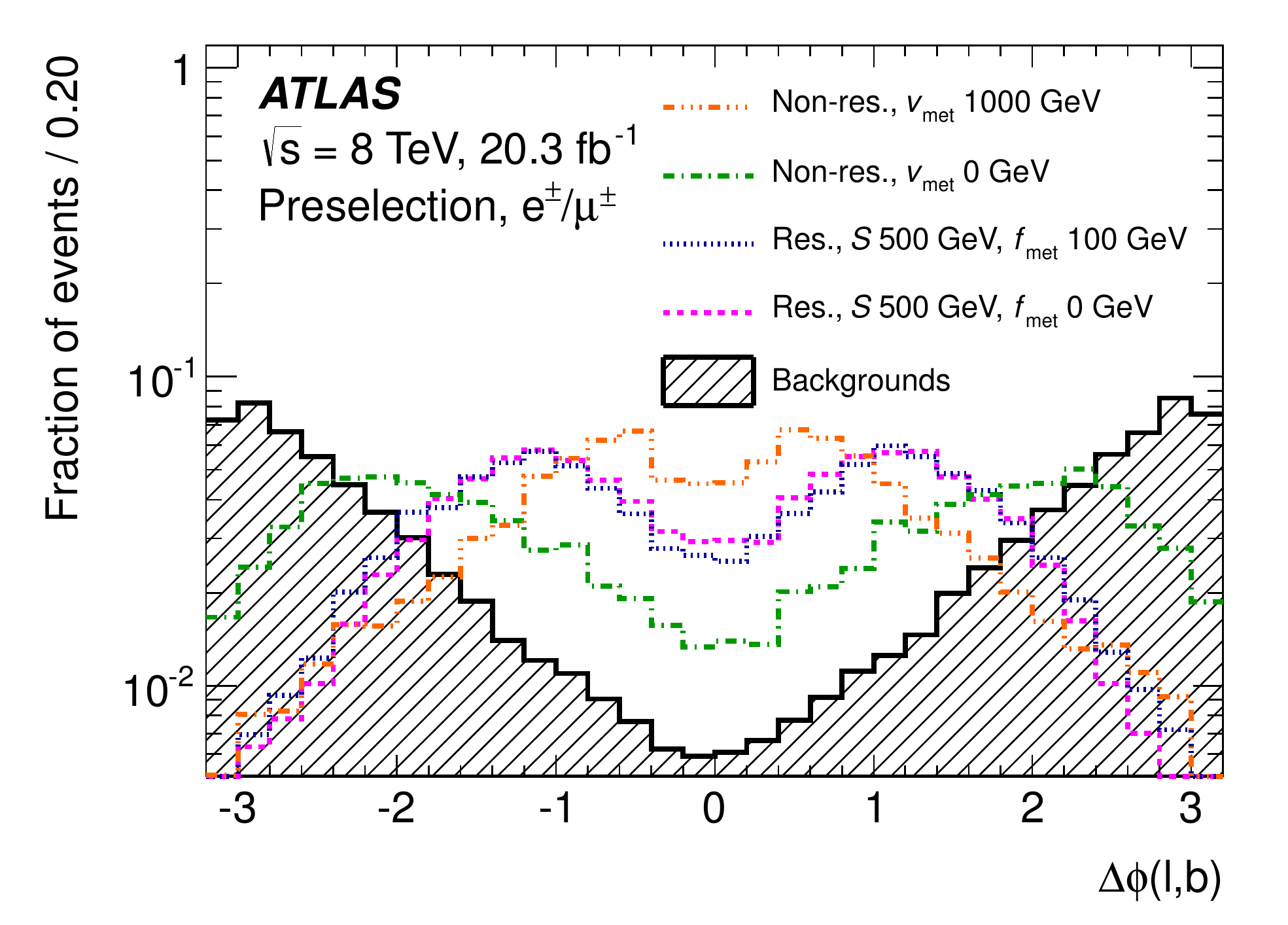}}
\caption
{
Distributions normalised to unity of \subref{subfig:BkgSig_MtW}~$\MTW$ and of \subref{subfig:BkgSig_DPhi}~$\deltaphilb$
for events satisfying the pre-selection defined in the text.
The expected distributions for the resonant model with $m(S)=500$~GeV are shown for the $\mfmet=0$~GeV and $\mfmet=100$~GeV
hypotheses, as well as for the non-resonant model for the $\mvmet=0$~GeV and $\mvmet=1000$~GeV hypotheses.
All distributions are compared to the expected distribution for the backgrounds.
For the $\MTW$ distributions, the last bin includes overflows.
}
\label{fig:BkgSig}
\end{figure}

Figure~\ref{fig:BkgSig} shows the expected distributions of \MTW\ and of the difference in azimuth
between the lepton and the $b$-tagged jet \deltaphilb\ for signal and background events.
The signal is prominent for high \MTW\ and low \adeltaphilb\ values; therefore, the event selection is optimised
using criteria on these two kinematic variables, in addition to the preselection defined above.
One optimal selection is defined for each of the two models by calculating the expected excluded cross-section with the procedure
explained below, including the systematic uncertainties:
\begin{itemize}
  \item SRI (resonant model optimisation): $\MTW>\mathrm{210~GeV}$ and $|\deltaphilb|<1.2$;
  \item SRII (non-resonant model optimisation): $\MTW>\mathrm{250~GeV}$ and $|\deltaphilb|<1.4$.
\end{itemize}
The background modelling is validated in dedicated control regions, orthogonal to the two signal regions.

\section{Results and interpretation}

\begin{figure}[!htb]
\centering
\subfigure[\label{subfig:DataMCPlots_SRI}]{\includegraphics[width=0.4\textwidth]{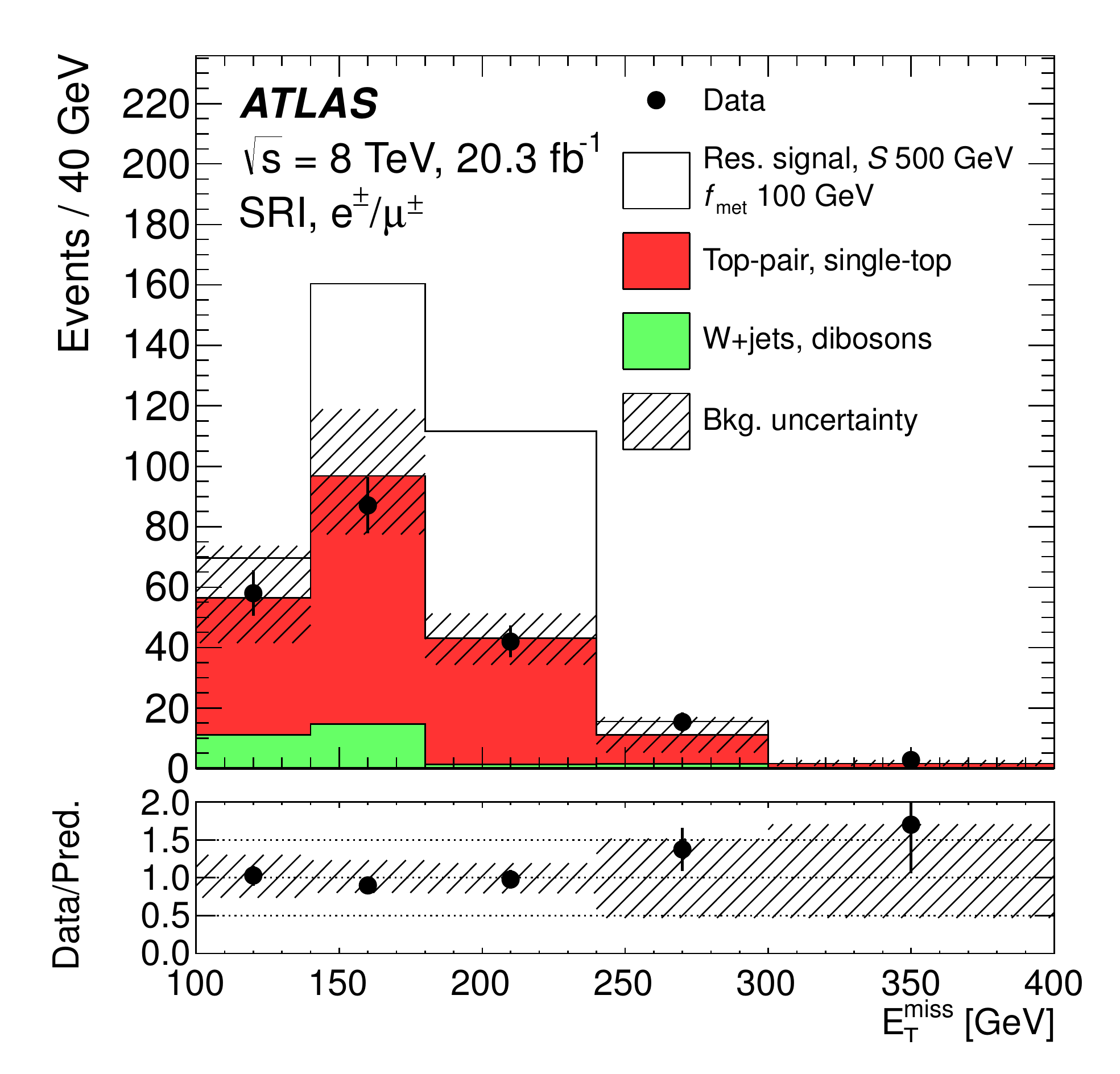}}
\subfigure[\label{subfig:DataMCPlots_SRII}]{\includegraphics[width=0.4\textwidth]{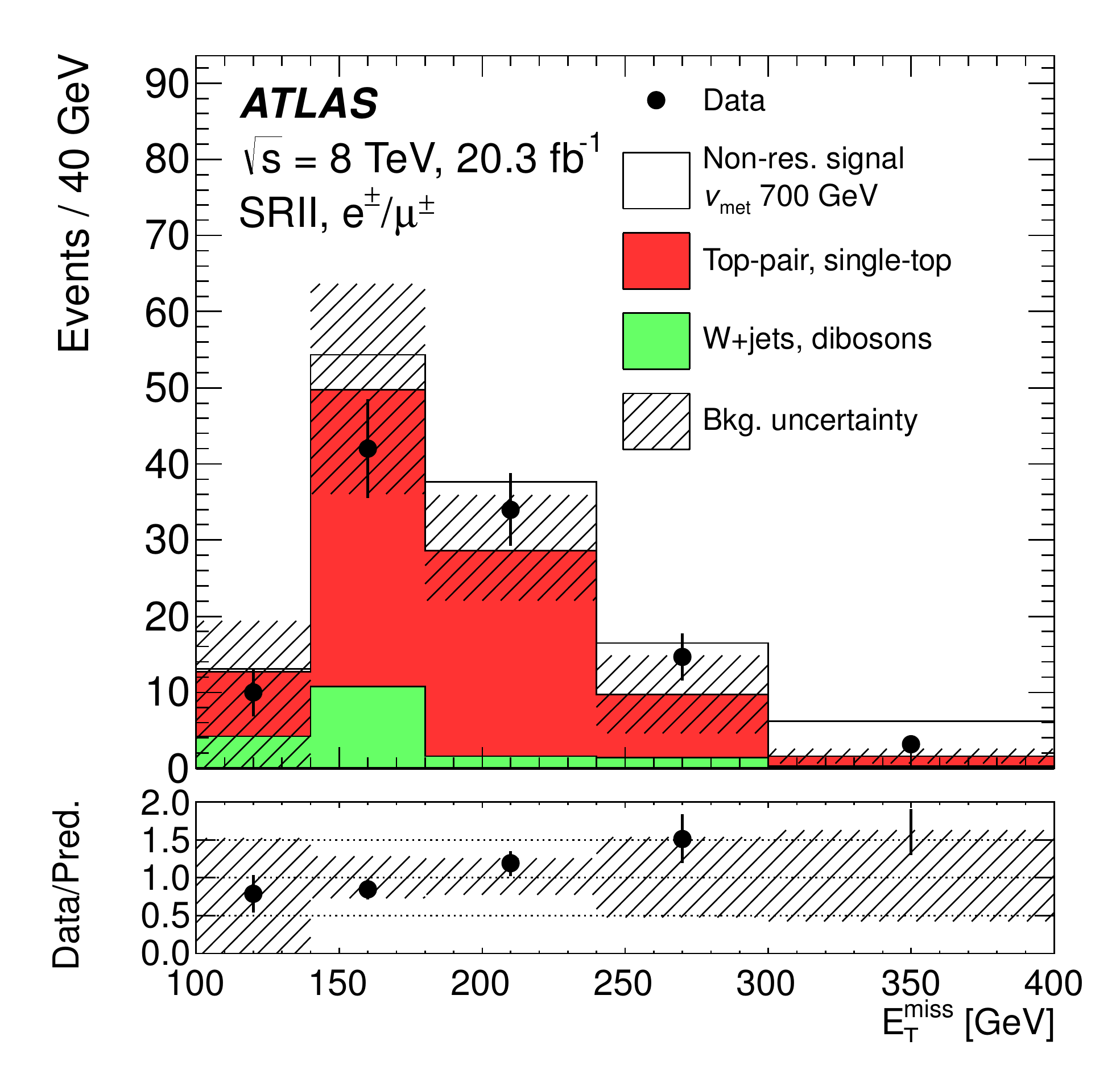}}
\caption
{
  Distributions of \MET\ in the \subref{subfig:DataMCPlots_SRI}~SRI and \subref{subfig:DataMCPlots_SRII}~SRII signal regions.
  The distributions observed in data, depicted with the points, are compared with the predicted background contributions,
  shown stacked together with the expected resonant (non-resonant) signal contribution for the $\mfmet=100$~GeV and $m(S)=500$~GeV
  ($\mvmet=700$~GeV) hypotheses.
  The expected backgrounds are normalised to their theoretical cross-sections, and the expected resonant (non-resonant) signal
  is normalised to the theoretical cross-section corresponding to $\ares=0.2$ ($\anonres=0.2$).
  The error bands on the expected backgrounds correspond to the uncertainties due to all systematic sources added in quadrature.
  The first (last) bin includes underflows (overflows).
  The ratios of the observed distributions to the predicted background distributions are shown in the lower frame.
}
\label{fig:DataMCPlots}
\end{figure}

Figure~\ref{fig:DataMCPlots} shows the \MET\ distributions in the SRI and SRII regions.
No significant deviation from the predicted background contribution is observed.
Therefore, 95\% confidence level (CL) upper limits on the signal production cross-sections are set using the $CL_s$ procedure.

\begin{figure}[!htb]
\centering
\subfigure[\label{subfig:XsecLimitRes}]{\includegraphics[width=0.42\textwidth]{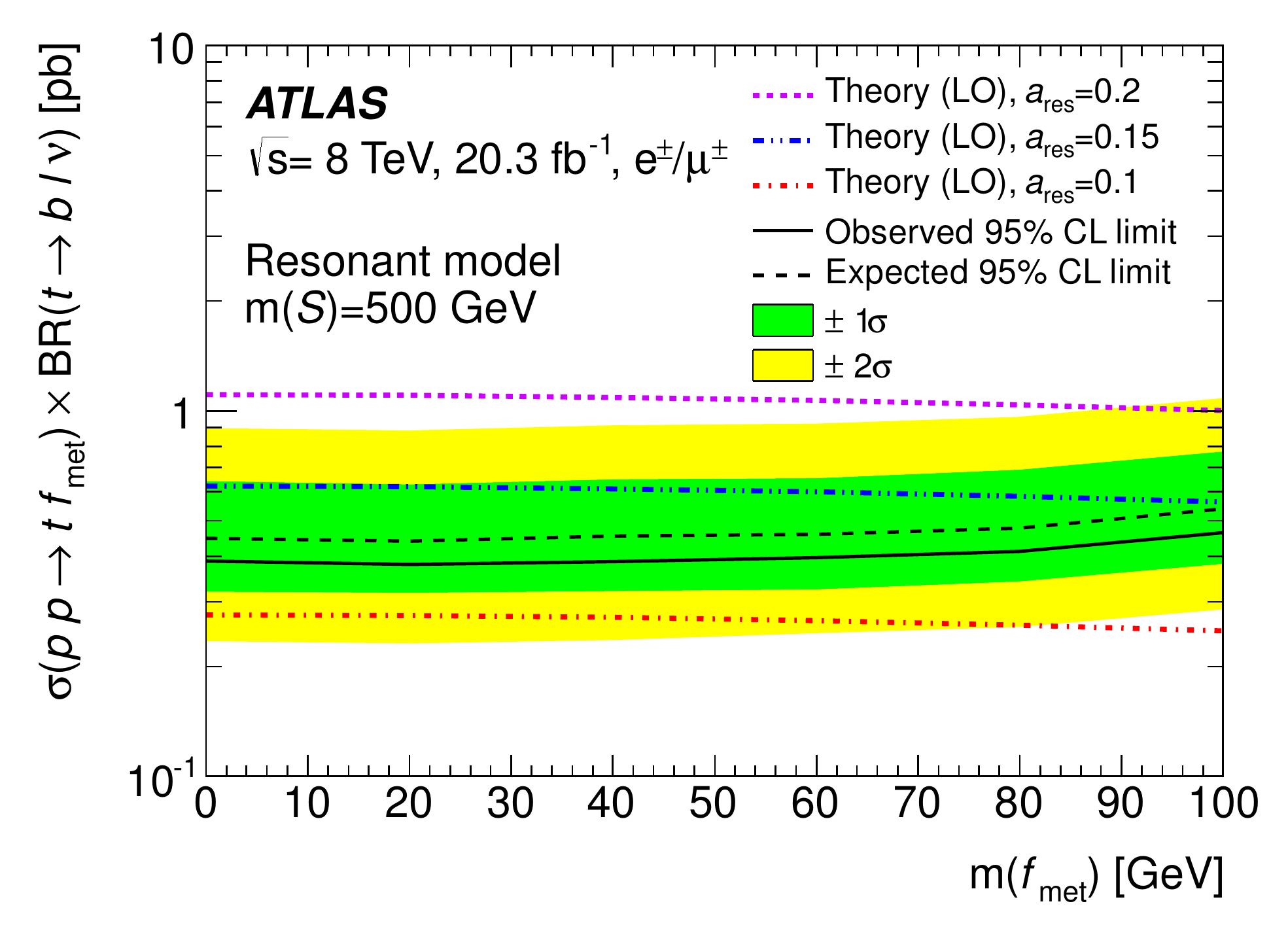}}
\subfigure[\label{subfig:XsecLimitNonRes}]{\includegraphics[width=0.42\textwidth]{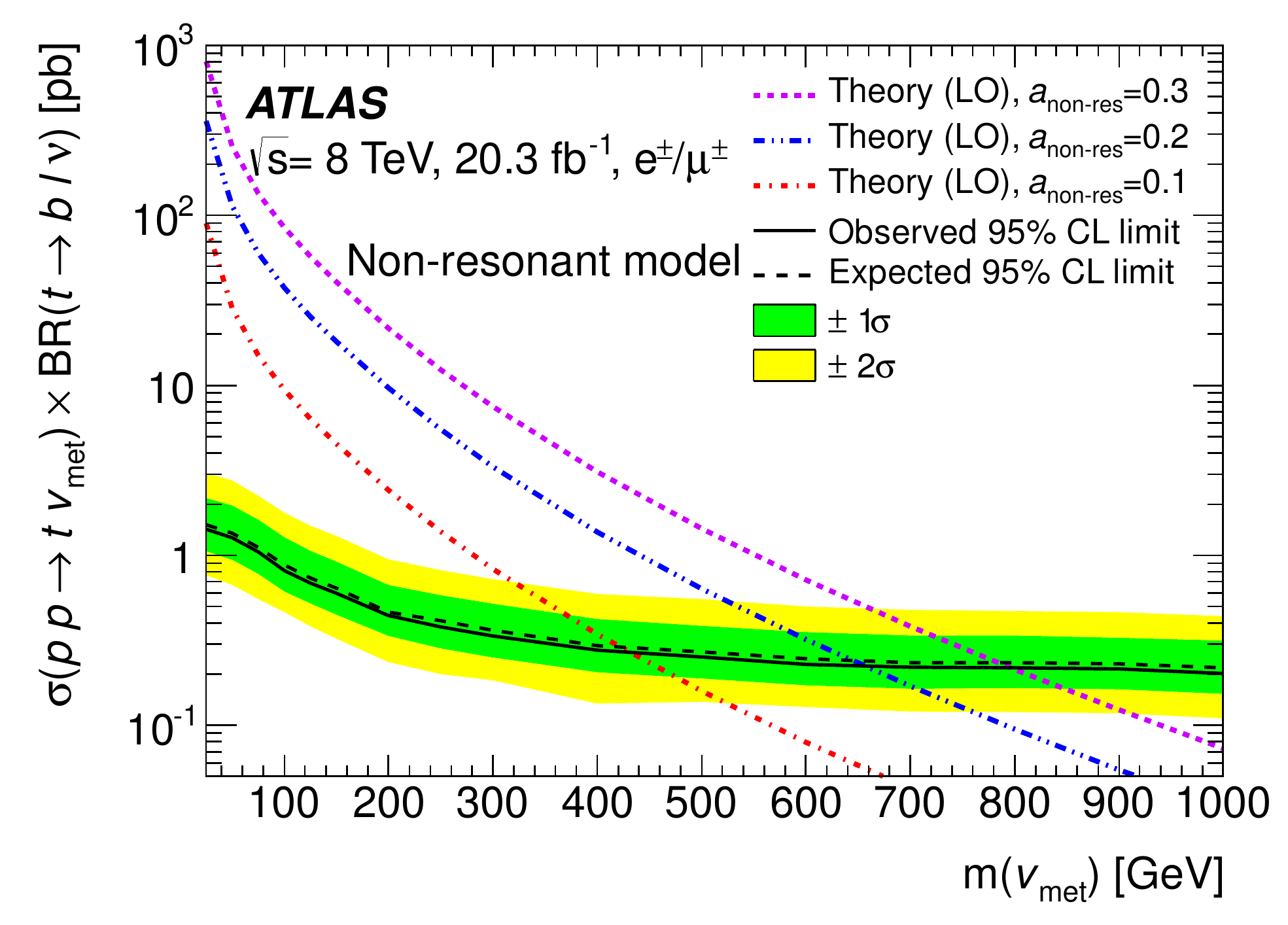}}
\caption
{
  Observed and expected limits on the cross section times branching ratio \subref{subfig:XsecLimitRes}~for the resonant
  model with $m(S)=500$~GeV and \subref{subfig:XsecLimitNonRes}~for the non-resonant model, as a function of the mass of $\fmet$ and $\vmet$, respectively.
}
\label{fig:XsecLimit}
\end{figure}

\begin{figure}[!htb]
\centering
\subfigure[\label{subfig:CoupLimitRes}]{\includegraphics[width=0.42\textwidth]{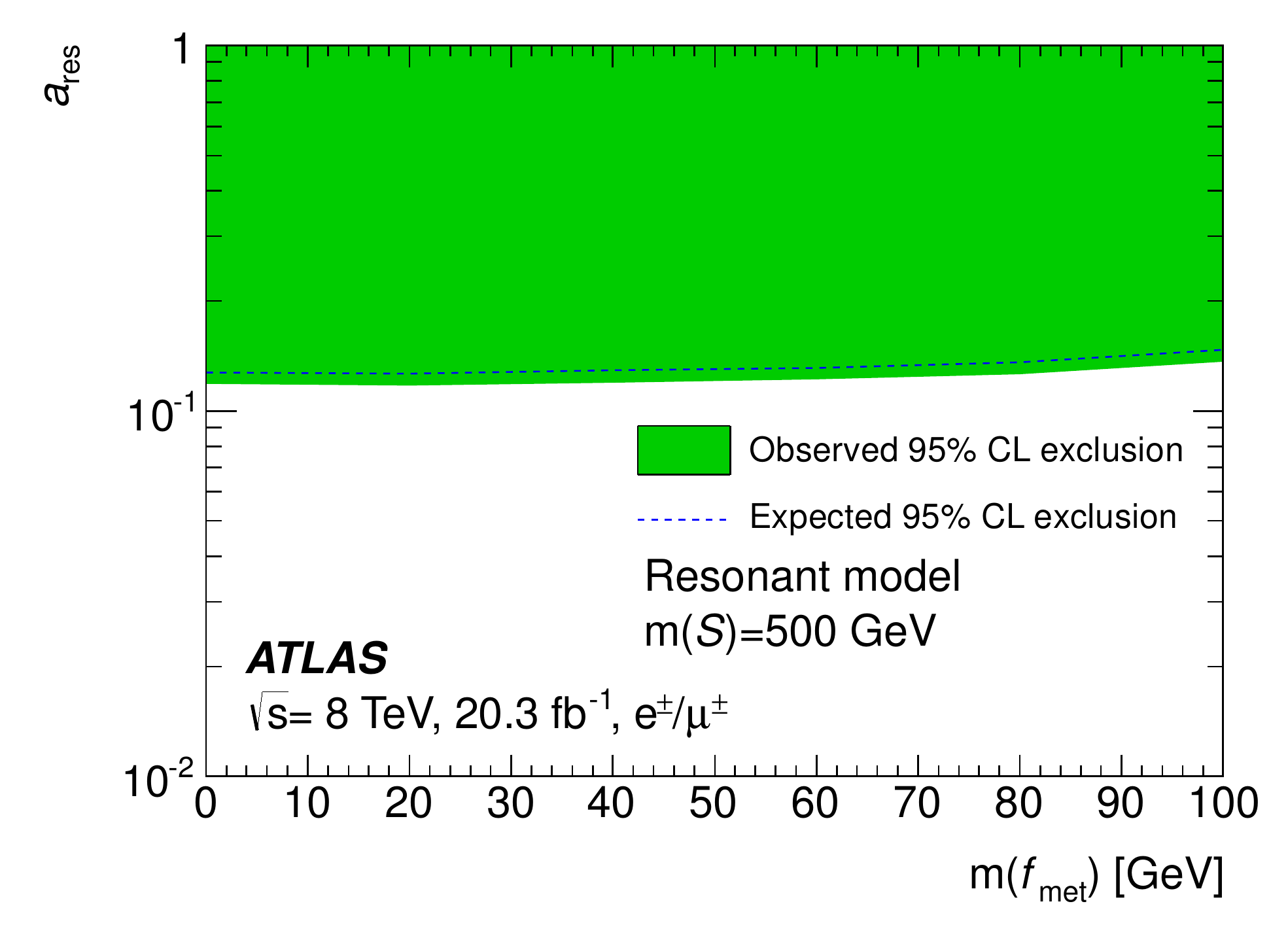}}
\subfigure[\label{subfig:CoupLimitNonRes}]{\includegraphics[width=0.42\textwidth]{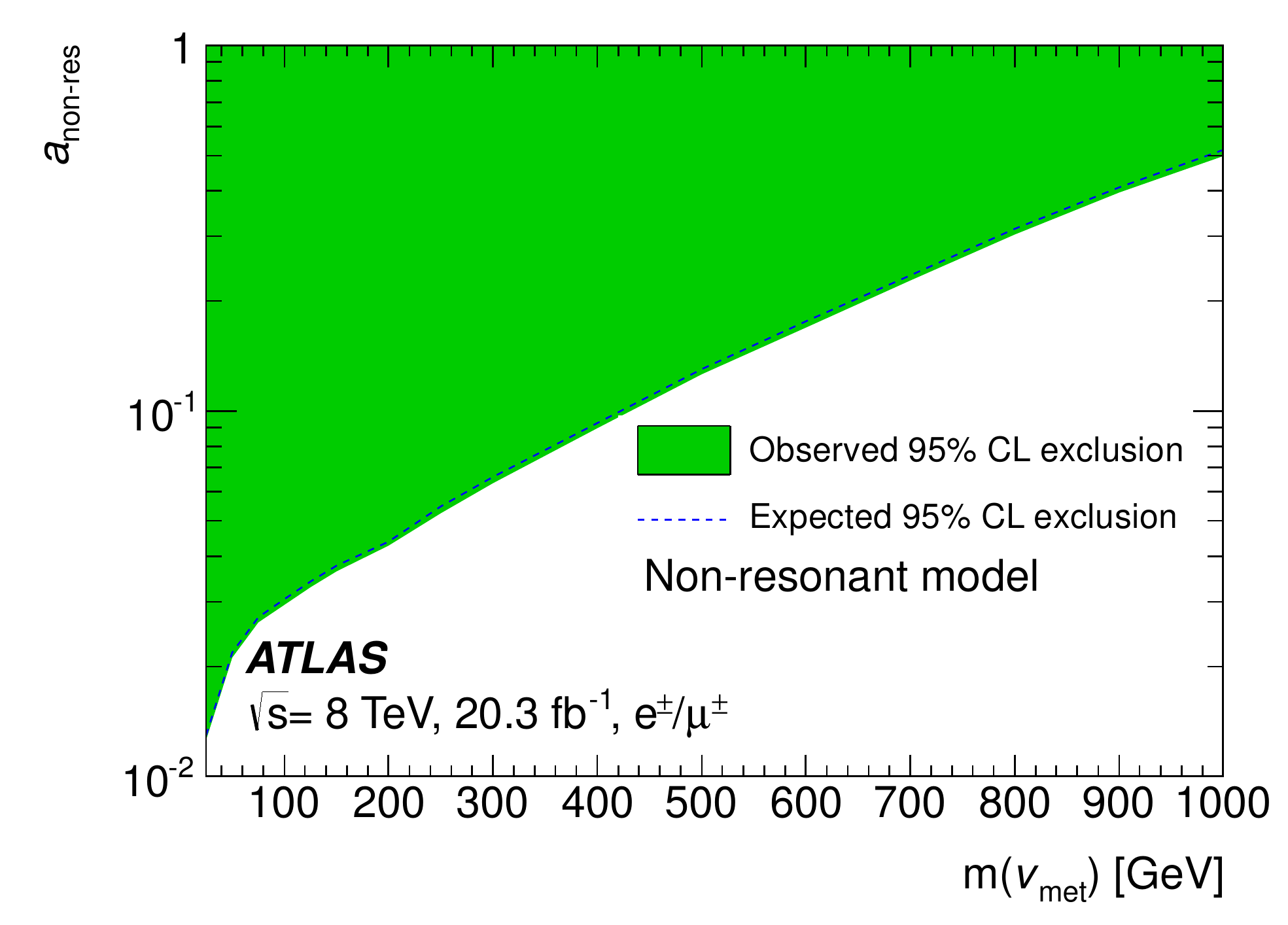}}
\caption
{
  Observed and expected excluded coupling strengths \subref{subfig:CoupLimitRes}~for the resonant model with $m(S)=500$~GeV
  and \subref{subfig:CoupLimitNonRes}~for the non-resonant model, as a function of the mass of $\fmet$ and $\vmet$, respectively.
}
\label{fig:CoupLimit}
\end{figure}

Figure~\ref{fig:XsecLimit} shows the limits on the signal production cross-sections multiplied by the branching ratio of the top decay mode,
for the resonant (non-resonant) model, as a function of the mass of $\fmet$ ($\vmet$).
In the resonant case, cross-sections corresponding to a coupling strength $\ares=0.2$ are excluded in the whole mass range.
In the non-resonant case, cross-sections corresponding to $\anonres=0.1$ are excluded up to $\mvmet=432$~GeV.
The limits on the cross-sections are used to calculate the limits on the coupling strengths shown in Figure~\ref{fig:CoupLimit}.

\FloatBarrier

\section*{References}

\bibliographystyle{atlasBibStyleWithTitle}
\bibliography{ATL-PHYS-PROC-2014-279}

\end{document}